\documentclass[aps,reprint,pra,longbibliography,nofootinbib]{revtex4-2}
\usepackage{amssymb}
\usepackage{amsmath}
\usepackage{dcolumn}
\usepackage{bm}
\usepackage{graphicx}
\usepackage{mathrsfs}
\usepackage[colorlinks,linkcolor=blue,anchorcolor=blue,citecolor=blue,urlcolor=blue]{hyperref}
\usepackage{color}
\usepackage{hyperref}

\begin{document}

\title{High-Performance Quantum Transduction with Correlated Noise}
\author{Yu-Bo Hou}
\thanks{Yu-Bo Hou and Xiaoan Ai contributed equally to this work.}
\author{Xiaoan Ai}
\thanks{Yu-Bo Hou and Xiaoan Ai contributed equally to this work.}
\author{Pengbo Li}
\author{Changchun Zhong}
\email{zhong.changchun@xjtu.edu.cn}

\address{
MOE Key Laboratory for Non-equilibrium Synthesis
and Modulation of Condensed Matter, Shaanxi Province Key Laboratory
of Quantum Information and Quantum Optoelectronic Devices, School of
Physics, Xi’an Jiaotong University, Xi’an 710049, China}

\begin{abstract}
Quantum transduction, which coherently converts quantum states between
microwave and optical frequency domains, is a key technology for
hybrid quantum architectures. Its performance, however, is fundamentally limited by
thermal noise. Direct quantum transduction is particularly susceptible to
noise and often fails to achieve positive quantum capacity.
Entanglement-based quantum transduction, which realizes state conversion through quantum teleportation assisted by microwave-optical entanglement, is intrinsically more robust against thermal noise. However, generating sufficiently strong entanglement in a realistic thermal environment remains a major challenge. In this paper, we
exploit correlated noise as a resource for quantum transduction. For
direct quantum transduction, it is shown that the noise correlations give rise to controllable interference terms that substantially suppress the effective channel noise.
For entanglement-based quantum transduction, the same correlations enhance the generation of microwave-optical entanglement, thereby improving the fidelity of teleportation-based conversion. As a
result, both transduction protocols exhibit broad regions of positive quantum capacity over experimentally relevant ranges of cooperativity. We further discuss a possible physical mechanism for engineering the required noise correlations, providing theoretical guidance for experimental
implementations. These results suggest that correlated noise can substantially relax the stringent cryogenic requirements for microwave-optical quantum transduction and facilitate the realization of practical hybrid quantum networks.
\end{abstract}

\maketitle

\section{Introduction}

Quantum transduction implements coherent transfer of quantum information
between distinct physical platforms, typically between microwave and optical
photons \cite%
{perspectivesonquantum,developmentofquantum,coherentconversionbetween}. It
is a key technology for large-scale distributed quantum networks because it
permits interfacing specialized local quantum nodes with long-distance
quantum communication channels \cite{quantumstatetransfer,thequantuminternet}%
. In particular, superconducting circuits offer
fast, high-fidelity quantum information processing and are therefore natural
candidates for quantum nodes \cite%
{circuitquantumelectrodynamics,wiringupquantum,quantuminformationprocessing}%
. However, they lack intrinsic optical transitions. Optical photons, by
contrast, propagate with low loss in fiber and preserve coherence over long
distances at ambient conditions, making them the preferred carriers for
interconnects \cite%
{violationofbell,satellite-basedentanglement,quantumteleportationover,anintegratedspace-to-ground}%
. Microwave-optical (MO) quantum transduction, which coherently converts
quantum information between the two disparate frequency domains, provides
the essential link between superconducting nodes and optical quantum
channels in quantum network architectures.

All quantum transduction processes can be rigorously described as quantum
channels, and only channels with positive quantum capacity can faithfully
transmit encoded quantum information. This basic requirement imposes stringent
constraints on a MO transducer, which must combine sufficiently high channel
transmissivity with a low level of added noise \cite%
{gaussianquantuminformation}. In practice this means maintaining a strong
coherent coupling strength while simultaneously suppressing thermal noise,
which is quite demanding \cite%
{cavitypiezo-mechanicsfor,cavityelectro-opticcircuit,proposalforheralded}.
Currently, the realization of a high-quality quantum transducer with the
state-of-the-art technology remains a challenge. Traditional direct quantum transduction (DQT) relies on beam-splitter interactions to coherently exchange excitations between microwave and optical modes \cite{bidirectionalandefficient,superconductingqubitto,bi-directionalconversionbetween}. In practice, however, limited conversion efficiency and excess noise often preclude a positive quantum capacity. Various strategies have therefore been explored to improve transduction performance, such as adaptive feedforward control \cite%
{harnessingelectro-opticcorrelations,quantumtransductionwith}, single-mode
squeezing \cite{quantumtransductionis} and entanglement-assisted enhanced
transduction \cite%
{overcomingthefundamental,intrabandentanglement--assistedcavity}.

Entanglement-based quantum transduction (EQT) provides an alternative
paradigm to DQT. In EQT, an unknown microwave or optical input state, which
carries the encoded quantum information, is transferred through quantum
teleportation with the help of pre-shared MO entanglement \cite{reversibleoptical-to-microwavequantum,proposalforheralded,deterministicmicrowave-Opticaltransduction}. By avoiding direct transmission of the input state through the noisy converter, this architecture is generally more robust against thermal noise than DQT \cite{gaussianquantuminformation,reversecoherentinformation,microwaveandoptical}. Its performance, however, is ultimately limited by the quality of the shared entanglement, which determines the effective transduction channel noise and, consequently, the achievable quantum capacity. Generating high-quality MO entanglement requires sufficiently strong coherent interactions while simultaneously suppressing thermal noise, posing a major challenge for current quantum transduction platforms \cite{reversibleoptical-to-microwavequantum,robustphotonentanglement,electro-opticentanglementsource}.

The detrimental effects of thermal noise on both DQT and EQT motivate approaches that go beyond simply reducing the thermal occupation of individual modes. Notably, correlated noise has recently emerged as a useful resource in a variety of quantum information tasks, including quantum sensing, feedforward correction, noise cancellation and the reactivation of quantum communication \cite{harnessingelectro-opticcorrelations,quantumtransductionwith,entanglement-enhancedsensingin,entanglement-enhancedoptomechanicalsensing,protectingquantuminformation,quantumbackactionand,mid-circuitcorrectionof,environment-assistedbosonicquantum,surpassingspectatorqubits}. In particular, Ref. \cite{correlatednoisecan} showed that, beyond conventional cooling techniques, engineered noise correlations among different modes can significantly improve the performance of DQT at elevated temperatures by introducing destructive interference that effectively suppresses the transduction channel noise.

In this work, we further investigate the role of noise correlations in quantum transduction. We first revisit the correlated noise model for DQT introduced in Ref. \cite{correlatednoisecan}. We then extend this framework to EQT and systematically examine how noise correlations can be exploited to enhance its performance. Finally, we discuss a potential way of engineering the required noise correlations. Specifically, by appropriately tuning the system parameters, the correlated noise contributions can be made to interfere destructively at the DQT output \cite{correlatednoisecan}. The resulting suppression of the effective channel noise substantially enhances the quantum capacity of DQT and relaxes the stringent requirements for achieving a positive quantum capacity. For EQT, we show that correlated noise can enhance the generation of microwave-optical entanglement, leading to improved transduction performance and positive quantum capacity over a broad region of the system parameter space. In addition, using the parameter set adopted in our numerical analysis, we demonstrate that the proposed correlation-engineering mechanism simultaneously improves the performance of both DQT and EQT. These results highlight noise correlations as a useful resource for quantum transduction, offering a route toward more robust interconnects for distributed quantum architectures.

\section{Dynamics of Electro-optomechanical System}

\begin{figure}[tbh]
\centering
\includegraphics[width=\columnwidth]{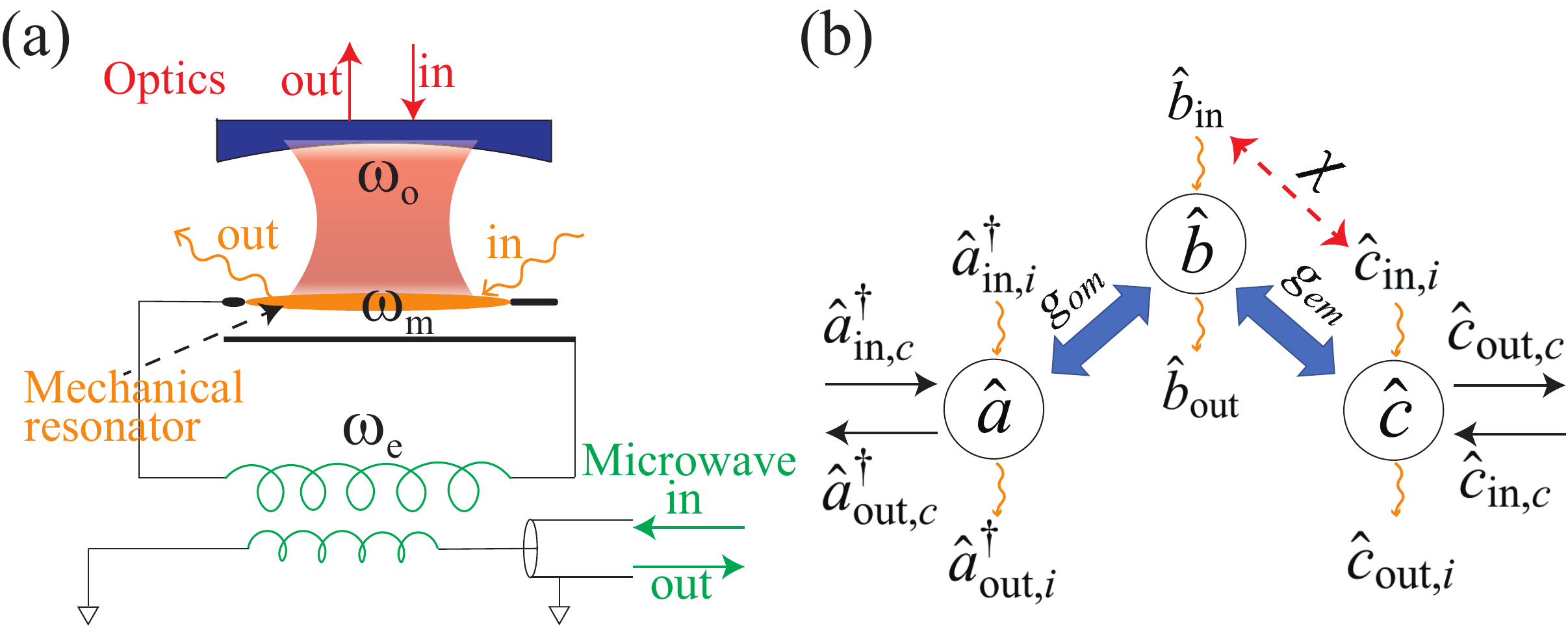}
\caption{(a) Schematic figure of an EOM system. (b) Illustration for the
input-output relations with correlated bath noise between the
microwave and mechanical modes. $\hat{a}$, $\hat{b}$ and $\hat{c}$ are
denoted as the annihilation operators of the optical, mechanical and
microwave modes, respectively, with the corresponding resonant frequencies $%
\protect\omega _{\mathrm{o}}$, $\protect\omega _{\mathrm{m}}$ and $\protect%
\omega _{\mathrm{e}}$. In plot (b), the black straight arrows and orange
wavy arrows denote the input and output fields associated with the coupling
and intrinsic loss ports, respectively, whereas the red dashed double-headed
arrows indicate correlations between the mechanical and microwave noise
fields.}
\label{f0}
\end{figure}

In this work, we adopt an electro-optomechanical (EOM) system for demonstration without loss of generality, as shown
in Fig. \ref{f0}(a). The system Hamiltonian is given by%
\begin{eqnarray}
\frac{\hat{H}}{\hbar } &=&\omega _{\mathrm{o}}\hat{a}^{\dag }\hat{a}+\omega
_{\mathrm{m}}\hat{b}^{\dag }\hat{b}+\omega _{\mathrm{e}}\hat{c}^{\dag }\hat{c%
}  \notag \\
&&+g_{\mathrm{o}}(\hat{b}+\hat{b}^{\dag })\hat{a}^{\dag }\hat{a}+g_{\mathrm{e}}(\hat{b}+\hat{b}^{\dag })\hat{c}^{\dag }\hat{c}\text{.}  \label{e1}
\end{eqnarray}%
Here, $\hat{a}$, $\hat{b}$ and $\hat{c}$ are denoted as
the annihilation operators of the optical, mechanical and microwave modes,
respectively, with the corresponding resonant frequencies $\omega _{\mathrm{o%
}}$, $\omega _{\mathrm{m}}$ and $\omega _{\mathrm{e}}$. $g_{\mathrm{o}}$ ($g_{\mathrm{e}}$) is the coupling rate between
the optical (microwave) mode and the mechanical mode. Since the bare single-photon coupling is typically weak, the EOM system is therefore operated in the
standard linearized regime under strong coherent pumping, with the pump
detunings appropriately chosen to activate the desired effective
interactions, e.g., beam-splitter or two-mode-squeezing processes \cite{cavityoptomechanics,reversibleoptical-to-microwavequantum,bidirectionalandefficient,nanomechanicalcouplingbetween,entanglingopticaland,opticaldetectionof}.

The EOM device acts as a linear Gaussian processor that maps Gaussian input fields to Gaussian output fields. The optical (electrical) mode is interfaced to the external circuitry via an
optical fiber (a microwave transmission line), which carries the input and
output fields with the external coupling rate denoted by $\kappa _{\mathrm{%
o,c}}$ ($\kappa _{\mathrm{e,c}}$). In practice, these modes are also intrinsically coupled to their thermal bath, whose loss rates are described by $\kappa _{\mathrm{o,i}}$, $\kappa
_{\mathrm{e,i}}$ and $\kappa _{\mathrm{m}}$ for the optical, electrical and
mechanical modes, respectively. The corresponding thermal noise is scattered into the output fields and can substantially degrade the fidelity of the converted state.
In conventional models, the thermal noise acting on different modes is assumed to be independent because it originates from distinct physical processes. However, this assumption
is not fundamental. If correlations exist among the noise sources, they can be exploited by appropriately tuning experimentally accessible system parameters so that their contributions to the output field interfere destructively, thereby partially canceling the effective thermal noise.

\section{Direct Quantum Transduction with Correlated Noise}

We now review the main idea of DQT under the correlated noise model \cite%
{correlatednoisecan}. In this stage, the optical (microwave) mode is driven by a strong
coherent pump with frequency $\omega _{\mathrm{1}}$ ($\omega _{\mathrm{2}}$) and detuning $\Delta _{\mathrm{1}}=\omega _{\mathrm{1}}-\omega _{\mathrm{o}}$ ($\Delta _{\mathrm{2}}=\omega _{\mathrm{2}}-\omega _{\mathrm{e}}$). When the red
sideband condition $\Delta _{\mathrm{1}}=\Delta _{\mathrm{2}}=-\omega _{\mathrm{m}}$ is
satisfied, linearization followed by the rotating-wave approximation (RWA) reduces both the optomechanical and electromechanical interactions to beam-splitter processes. The effective Hamiltonian in the rotating frame
is then given by (Hereafter, all effective Hamiltonians are written in the
interaction picture with respect to the free Hamiltonian.)%
\begin{equation}
\frac{\hat{H}_{1}}{\hbar }=g_{\mathrm{om}}(\hat{a}^{\dag }\hat{b}+\hat{a}%
\hat{b}^{\dag })+(g_{\mathrm{em}}\hat{b}^{\dag }\hat{c}+g_{\mathrm{em}%
}^{\ast }\hat{b}\hat{c}^{\dag })\text{,}  \label{e13}
\end{equation}%
where $g_{\mathrm{om}}$ and $g_{\mathrm{om}}$ denote the pump-enhanced coupling rates. Note that
$g_{\mathrm{em}}=|g_{\mathrm{em}}|e^{i\phi }$ is kept complex because the
coupling phase $\phi $ plays an important role in the correlated noise interference discussed below. To
parameterize the strengths of the two linearized interactions, we introduce
the cooperativities $C_{\mathrm{em}}=4|g_{\mathrm{em}}|^{2}/\kappa _{\mathrm{%
m}}\kappa _{\mathrm{e}}$ and $C_{\mathrm{om}}=4g_{\mathrm{om}}{}^{2}/\kappa
_{\mathrm{o}}\kappa _{\mathrm{e}}$, with the the total cavity linewidths of
the optical and microwave modes given by $\kappa _{\mathrm{o}}=\kappa _{\mathrm{o,c}%
}+\kappa _{\mathrm{o,i}}$ and $\kappa _{\mathrm{e}}=\kappa _{\mathrm{e,c}%
}+\kappa _{\mathrm{e,i}}$, respectively. Moreover, the optical (microwave)
mode extraction ratio is defined as $\zeta _{\mathrm{o(e)}}=\kappa _{\mathrm{%
o(e),c}}/\kappa _{\mathrm{o(e)}}$.

In this configuration, MO conversion can be viewed as a cascaded mode-swap
process mediated by the two beam-splitter interactions. This process is
described quantitatively using the scattering picture framework, and the scattering relation can be solved in the narrow-bandwidth limit ($\omega =0$) as
\begin{equation}
\hat{a}_{\mathrm{out,c}}=\sqrt{\eta }\hat{c}_{\mathrm{in,c}}+\sqrt{1-\eta }%
\hat{e}\text{,}  \label{e21}
\end{equation}
with the operator $\hat{e}$ given by
\begin{equation}
\hat{e}=\frac{1}{\sqrt{1-\eta }}(S_{11}\hat{a}_{\mathrm{in,c}}+S_{12}\hat{a}%
_{\mathrm{in,i}}+S_{13}\hat{b}_{\mathrm{in}}+S_{15}\hat{c}_{\mathrm{in,i}})%
\text{,}  \label{e22}
\end{equation}
where $\eta =4C_{\mathrm{om}}C_{\mathrm{em}}\zeta _{\mathrm{o}}\zeta _{\mathrm{e}}/(1+C_{\mathrm{om}}+C_{\mathrm{em}})^{2}$ is the transmissivity and $S_{ij}$ denotes the corresponding element of the system scattering matrix \cite{correlatednoisecan}. Here, the lower indexes \textquotedblleft
in\textquotedblright\ and \textquotedblleft out\textquotedblright\ are used
to indicate the fluctuation operators of the input and output traveling
modes, while \textquotedblleft c\textquotedblright\ and \textquotedblleft
i\textquotedblright\ represent the coupling and intrinsic loss ports,
respectively.

Based on this, the microwave-to-optical
conversion defines a bosonic thermal loss channel with conversion efficiency $\eta $ and thermal
noise $n_{e}=\langle \hat{e}^{\dag }\hat{e}\rangle $, and we denote it as $%
\mathcal{N}(\eta ,n_{e})$. In the quadrature representation, this Gaussian
channel maps an input Gaussian state with covariance matrix $\mathbf{V}$
into $\mathbf{TVT}^{\mathrm{T}}+\mathbf{N}$ with $\mathbf{T}=\sqrt{\eta }%
\mathbf{I}_{2}$ and $\mathbf{N}=(1-\eta )(2n_{e}+1)\mathbf{I}_{2}$. Here, we
assumed that the microwave and mechanical modes are intrinsically coupled to
the thermal bath with the same temperature $\mathcal{T}$. Consequently, the
thermal photon excitation number is given by $n_{\mathrm{th}}=\langle \hat{b}%
_{\mathrm{in}}^{\dag }\hat{b}_{\mathrm{in}}\rangle = \langle \hat{c}_{%
\mathrm{in,i}}^{\dag }\hat{c}_{\mathrm{in,i}}\rangle = \lbrack \exp
(\hbar \omega _{\mathrm{e(m)}}/k_{\mathrm{B}}\mathcal{T})-1]^{-1}$, where $%
k_{\mathrm{B}}$ is the Boltzmann constant. For the large frequency of the
optical mode (usually with $\omega _{\mathrm{o}}>100$ THz), its thermal
photon number $\langle \hat{a}_{\mathrm{in,i}}^{\dag }\hat{a}_{\mathrm{in,i}%
}\rangle $ is very small even at room temperature. So the optical thermal
occupation can be safely ignored.

To quantify the performance of the transduction channel, we use the quantum
capacity as the figure of merit for its ability to transmit quantum
information. For many quantum channels, however, the exact quantum capacity
is difficult to determine. A practical approach is therefore to use an
achievable lower bound. For the bosonic thermal loss channel $\mathcal{N}%
(\eta ,n_{e})$, a commonly used lower bound on the quantum capacity is given
by \cite{gaussianquantuminformation}%
\begin{equation}
Q_{\mathrm{LB}}^{\mathcal{N}}=\max \left\{ 0,\log _{2}\frac{\eta }{1-\eta }%
-g(n_{e})\right\} \text{,}  \label{e23}
\end{equation}%
where $g(x)\equiv (x+1)\log _{2}(x+1)-x\log _{2}x$ is the monotonically
increasing bosonic entropy function. This bound becomes tight for the pure
loss channel with $n_{e}=0$. In this limit, a positive quantum capacity,
which is the threshold for quantum information to be reliably transmitted,
requires $\eta >\frac{1}{2}$. For a general thermal loss channel with $%
n_{e}>0$, this imposes a more stringent requirement on the transmissivity $%
\eta $.

Here, we assume the noise operators are correlated as%
\begin{equation}
\langle \hat{c}_{\mathrm{in,i}}^{\dag }\hat{b}_{\mathrm{in}}\rangle \equiv
n_{\mathrm{th}}\cdot \chi \text{,} \label{ex}
\end{equation}%
where $\chi $ is the measure of the degree of correlations with $0\leq |\chi
|\leq 1$. Taking $\chi $ to be a positive real number, we can write the effective channel noise as%
\begin{equation}
n_{e}=\frac{1}{1-\eta }(|S_{13}|^{2}+|S_{15}|^{2}+S_{13}^{\ast }S_{15}\chi
+S_{15}^{\ast }S_{13}\chi )n_{\mathrm{th}}\text{.}
\end{equation}%
By appropriately selecting the coupling phase $\phi $, the two additional
terms arising from the noise correlations%
\begin{equation}
S_{13}^{\ast }S_{15}+S_{15}^{\ast }S_{13}=\frac{8\sqrt{C_{\mathrm{em}%
}(1-\zeta _{\mathrm{e}})}C_{\mathrm{om}}\zeta _{\mathrm{o}}}{(1+C_{\mathrm{om%
}}+C_{\mathrm{em}})^{2}}\sin \phi \text{,}
\end{equation}
can be made negative for $\pi <\phi <2\pi $. That is, they can
interfere destructively with the independent noise contributions, thereby
effectively suppressing $n_{e}$. Consequently, the correlated noise model
significantly improves the transduction performance of DQT even at elevated
bath temperatures. Specifically, the marked reduction in $n_{e}$ leads to a
substantial increase in $Q_{\mathrm{LB}}^{\mathcal{N}}$ compared with the
independent noise model at the same bath temperature and greatly expands the
region with $Q_{\mathrm{LB}}^{\mathcal{N}}>0$ in the $C_{\mathrm{om}}-C_{%
\mathrm{em}}$ cooperativity space.

\section{Entanglement-Based Quantum Transduction with Correlated Noise}

\subsection{Generation of the microwave-optical entanglement resource}

In the previous section, we have reviewed that the correlated noise model
can effectively suppress the channel noise in DQT and reduce the requirement
for obtaining a positive quantum capacity. We now extend this idea to EQT. To this end, we first investigate how the correlated noise model affects the entanglement of the MO Gaussian state used in the teleportation protocol. For the
entanglement generation, we change the optical pump configuration, where the optical mode is driven by a strong coherent
pump with frequency $\omega _{\mathrm{3}}$ and detuning $\Delta _{\mathrm{3}%
}=\omega _{\mathrm{3}}-\omega _{\mathrm{o}}$. Under the blue sideband
condition $\Delta _{\mathrm{3}}=\omega _{\mathrm{m}}$, linearization
followed by the RWA reduces the optomechanical interaction to a two-mode
squeezing process. The resulting
effective Hamiltonian in the rotating frame is then given by%
\begin{equation}
\frac{\hat{H}_{2}}{\hbar }=g_{\mathrm{om}}(\hat{a}^{\dag }\hat{b}^{\dag }+%
\hat{a}\hat{b})+(g_{\mathrm{em}}\hat{b}^{\dag }\hat{c}+g_{\mathrm{em}}^{\ast
}\hat{b}\hat{c}^{\dag })\text{.} \label{e101}
\end{equation}%
As in the DQT case, we express the electromechanical coupling as $g_{\mathrm{%
em}}=|g_{\mathrm{em}}|e^{i\phi }$, where the phase $\phi $ is kept.

In this configuration, the MO entanglement is generated through the
combination of the two interactions in Hamiltonian (\ref{e101}). More specifically, as
a two-mode squeezed state is generated between the optical and mechanical
modes, the mechanical excitations are subsequently swapped into the
microwave mode via the electromechanical beam-splitter interaction, thereby
mediating an entangled MO output state. Since the dynamics is linear, the
output Gaussian state is fully determined by the scattering relation between
the input and output fields. Starting from $\hat{H}_{2}$, the
corresponding Heisenberg-Langevin equations and input-output relations are
written as%
\begin{equation}
\mathbf{\dot{a}}=\mathbf{Ma+Na}_{\mathrm{in}}\text{,}  \label{e26}
\end{equation}%
\begin{equation}
\mathbf{a}_{\mathrm{out}}=\mathbf{N}^{\mathrm{T}}\mathbf{a-a}_{\mathrm{in}}%
\text{,}  \label{e27}
\end{equation}%
with the vectors $\mathbf{a}=(\hat{a}^{\dag },\hat{b},\hat{c})^{\mathrm{T}}$%
, $\mathbf{a}_{\mathrm{in}}=(\hat{a}_{\mathrm{in,c}}^{\dag },\hat{a}_{%
\mathrm{in,i}}^{\dag },\hat{b}_{\mathrm{in}},\hat{c}_{\mathrm{in,c}},\hat{c}%
_{\mathrm{in,i}})^{\mathrm{T}}$ and $\mathbf{a}_{\mathrm{out}}=(\hat{a}_{%
\mathrm{out,c}}^{\dag },\hat{a}_{\mathrm{out,i}}^{\dag },\hat{b}_{\mathrm{out%
}},\hat{c}_{\mathrm{out,c}},\hat{c}_{\mathrm{out,i}})^{\mathrm{T}}$. Here,
the coefficient matrices are given by%
\begin{equation}
\mathbf{M}=\left(
\begin{array}{ccc}
-\frac{\kappa _{\mathrm{o}}}{2} & ig_{\mathrm{om}} & 0 \\
-ig_{\mathrm{om}} & -\frac{\kappa _{\mathrm{m}}}{2} & -ig_{\mathrm{em}} \\
0 & -ig_{\mathrm{em}}^{\ast } & -\frac{\kappa _{\mathrm{e}}}{2}%
\end{array}%
\right) \text{,}  \label{e28}
\end{equation}%
and
\begin{equation}
\mathbf{N}=\left(
\begin{array}{ccccc}
\sqrt{\kappa _{\mathrm{o,c}}} & \sqrt{\kappa _{\mathrm{o,i}}} & 0 & 0 & 0 \\
0 & 0 & \sqrt{\kappa _{\mathrm{m}}} & 0 & 0 \\
0 & 0 & 0 & \sqrt{\kappa _{\mathrm{e,c}}} & \sqrt{\kappa _{\mathrm{e,i}}}%
\end{array}%
\right) \text{.}  \label{e28}
\end{equation}%
Using the Fourier transform $\hat{o}(t)=(1/2\pi )\int_{-\infty }^{+\infty }%
\hat{o}(\omega )e^{-i\omega t}d\omega $, where $\hat{o}$ denotes an
arbitrary operator, the corresponding scattering relation in the
narrow-bandwidth limit is obtained as%
\begin{equation}
\mathbf{a}_{\mathrm{out}}=\mathbf{S}^{\prime }\cdot \mathbf{a}_{\mathrm{in}}%
\text{,}  \label{e29}
\end{equation}%
where the scattering matrix is given by $\mathbf{S}^{\prime }=\mathbf{N}^{%
\mathrm{T}}(-\mathbf{M)}^{-1}\mathbf{N}-\mathbf{I}_{5}$.

To fully characterize the entanglement of the MO two-mode output state, we
need to construct its covariance matrix. This requires transforming the
scattering matrix into the canonical quadrature representation. Applying the
relation%
\begin{equation}
\binom{\hat{q}}{\hat{p}}=\left(
\begin{array}{cc}
1 & 1 \\
-i & i%
\end{array}%
\right) \binom{\hat{a}}{\hat{a}^{\dag }}\text{,}  \label{e30}
\end{equation}%
the scattering relation (\ref{e29}) is converted into the
corresponding quadrature representation
\begin{equation}
\mathbf{x}_{\mathrm{out}}=\mathbf{S}\cdot \mathbf{x}_{\mathrm{in}}\text{,}
\label{e31}
\end{equation}%
where all the input (output) mode quadratures are collected into the vectors
$\mathbf{x}_{\mathrm{in}}$ ($\mathbf{x}_{\mathrm{out}}$) and $\mathbf{S}$ is
the corresponding symplectic map.

To determine the covariance matrix of the output MO Gaussian state in the
correlated noise model, we adopt the noise correlation setting as in
Ref. \cite{correlatednoisecan}, and the correlation
formula is written as%
\begin{equation}
\langle \hat{c}_{\mathrm{in,i}}^{\dag }\hat{b}_{\mathrm{in}}\rangle =\langle
\hat{b}_{\mathrm{in}}\hat{c}_{\mathrm{in,i}}^{\dag }\rangle \equiv n_{%
\mathrm{th}}\cdot \chi \text{,} \label{ex1}
\end{equation}%
as illustrated by the input-output schematic in Fig. \ref{f0}(b). If we
label the MO output state quadratures as $\mathbf{x}=\{\hat{q}_{\mathrm{o}},%
\hat{p}_{\mathrm{o}},\hat{q}_{\mathrm{e}},\hat{p}_{\mathrm{e}}\}^{\mathrm{T}%
} $, where the lower indexes \textquotedblleft o\textquotedblright\ and
\textquotedblleft e\textquotedblright\ respectively represent the optical
and microwave modes, we can obtain the corresponding covariance matrix $%
\mathbf{V}_{\mathrm{oe}}^{\mathrm{out}}$. According to Eq. (\ref{e31}), the
matrix elements can be calculated by $V_{ij}=1/2\langle \{\hat{x}%
_{i}-\langle \hat{x}_{i}\rangle ,\hat{x}_{j}-\langle \hat{x}_{j}\rangle
\}\rangle $. After applying local phase-space rotations, we can write down
the standard form of the MO covariance matrix%
\begin{equation}
\mathbf{V}_{\mathrm{oe}}=\left(
\begin{array}{cc}
\mathbf{V}_{\mathrm{A}} & \mathbf{V}_{\mathrm{C}} \\
\mathbf{V}_{\mathrm{C}}^{\mathrm{T}} & \mathbf{V}_{\mathrm{B}}%
\end{array}%
\right) \text{,}  \label{e32}
\end{equation}%
with the submatrices $\mathbf{V}_{\mathrm{A}}=u\mathbf{I}_{2}$, $\mathbf{V}_{%
\mathrm{B}}=v\mathbf{I}_{2}$ and $\mathbf{V}_{\mathrm{C}}=w\mathbf{Z}_{2}$,
where $\mathbf{Z}_{2}$ is the Pauli-z matrix. Here, the diagonal elements $u$
and $v$ represent the local quadrature noise spectra while the element $w$
indicates the quadrature correlations of the two modes, and their specific forms are given in Appendix \ref{aa}.

Notably, the parameters $u$, $v$ and $w$ defining $\mathbf{V}_{\mathrm{oe}}$
differ markedly between the correlated and independent noise models.
Intuitively, as in DQT, the correlated noise is transferred to the microwave
output through the electromechanical beam-splitter interaction in $\hat{H}%
_{2}$, thereby directly affecting the entanglement of the resulting MO
output state. Specifically, this effect is reflected in the explicit
dependence of the matrix elements in $\mathbf{V}_{\mathrm{oe}}$ on the
degree of noise correlation $\chi $ and the electromechanical coupling phase
$\phi $. The key point is that the phase $\phi $ not only controls whether the
correlated noise terms increase or reduce the local
quadrature variances $u$ and $v$, but also changes the intermode covariance $%
w$. Therefore, the quality of the generated MO entanglement is jointly
determined by their combined effects, and a proper choice of $\phi $ is
essential for improving the entanglement resource.

\begin{figure}[tbh]
\centering
\includegraphics[width=\columnwidth]{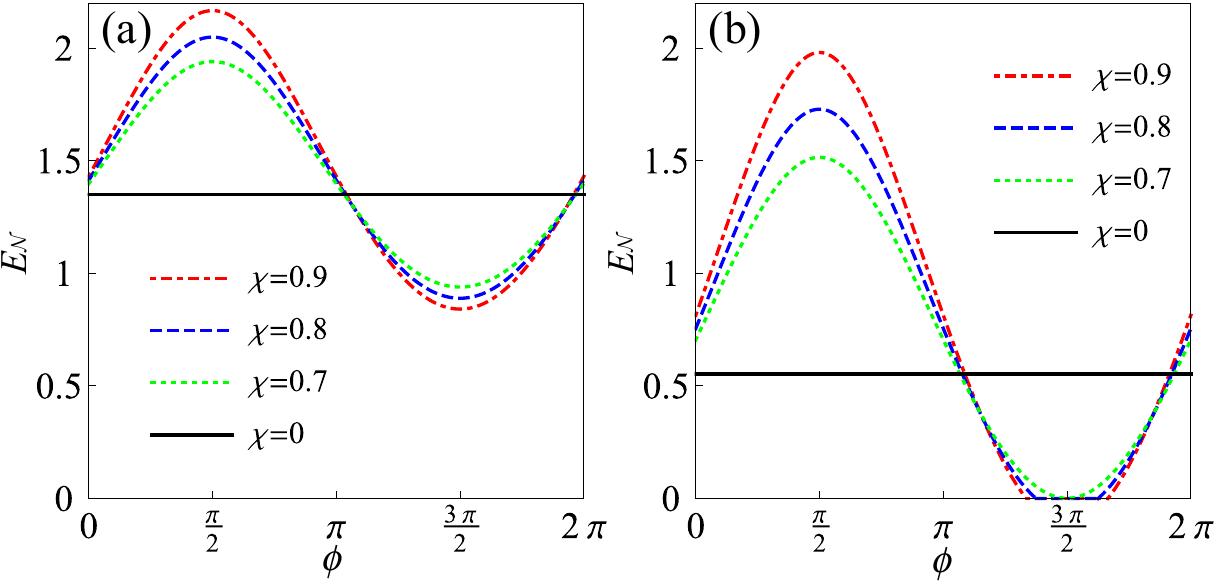}
\caption{Logarithmic negativity $E_{\mathcal{N}}$ of the MO output state
versus the electromechanical coupling phase $\protect\phi $ under different
degrees of noise correlation $\protect\chi =\{0,0.7,0.8,0.9\}$. Here, we
choose the frequency $\protect\omega _{\mathrm{m}}=\protect\omega _{\mathrm{e%
}}=10$ GHz, and $C_{\mathrm{om}}=C_{\mathrm{em}}=4$ and $\protect\zeta _{%
\mathrm{o}}=\protect\zeta _{\mathrm{e}}=0.9$ are used for both figures (a)
and (b). The thermal noise is chosen as (a) $n_{\mathrm{th}}=0.6$ and (b) $%
n_{\mathrm{th}}=1.6$.}
\label{f10}
\end{figure}

\begin{figure*}[tbh]
\centering \includegraphics[width=17.9cm]{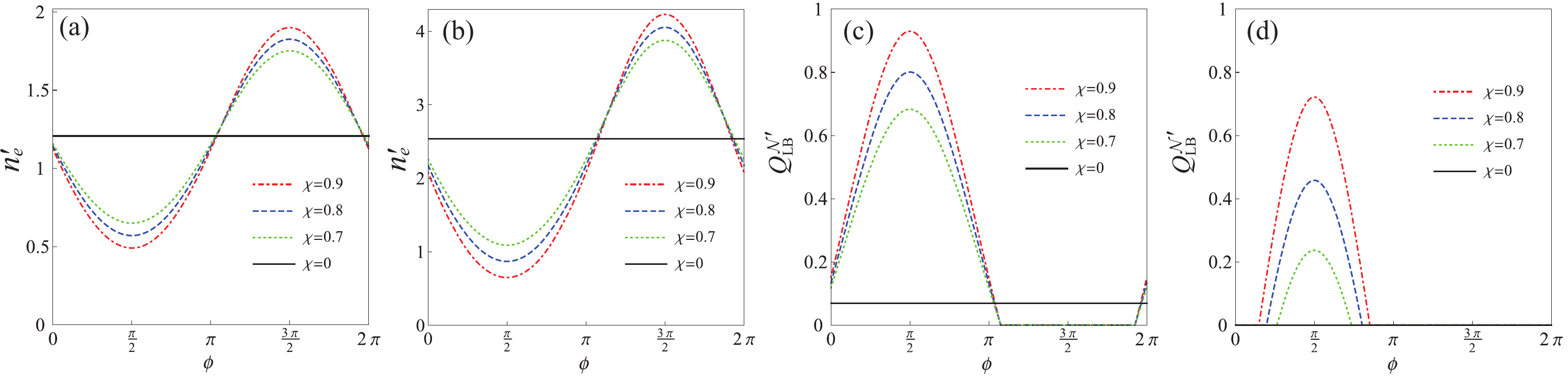}
\caption{Effective added noise $n_{e}^{\prime }$ in panels (a) and (b) and quantum
capacity lower bound $Q_{\mathrm{LB}}^{\mathcal{N}^{\prime }}$ in panels (c)
and (d) versus electromechanical coupling phase $\protect\phi $. The thermal noise is chosen as $n_{\mathrm{th}}=0.6$ in (a) and (c), and $n_{\mathrm{th}}=1.6$ in (b) and (d). The optimal gain constant $\protect\kappa $ was chosen
for each parameter set. The other parameters are chosen to be the same as in
Fig. \protect\ref{f10}.}
\label{f11}
\end{figure*}

To quantify the entanglement of the MO output state, we use the logarithmic
negativity $E_{\mathcal{N}}$ \cite{computablemeasureof}, which is a standard measure for two-mode
Gaussian entanglement and vanishes for separable two-mode Gaussian states (more details in Appendix \ref{aa}). Fig. \ref{f10}(a) shows $E_{\mathcal{N}}$ as a function of the coupling
phase $\phi $ under different degrees of noise correlation $\chi $. We see
that compared to the uncorrelated noise with $\chi =0$, larger $E_{\mathcal{N%
}}$ can be achieved when choosing an appropriate coupling phase, and the
optimal choice is $\phi =\pi /2$. Moreover, the advantages of correlated
noise will be more apparent if the bath temperature is higher, as
illustrated in Fig. \ref{f10}(b). It is worth noting that an inappropriate
choice of $\phi $ can reduce the entanglement to a level even lower than
that obtained in the independent noise model, as indicated by the curves
that fall below the black solid line.

\subsection{Teleportation-induced transduction channel}

In the correlated noise model we can prepare MO output states with enhanced
entanglement. These states can be used as resources for teleportation to
realize bidirectional quantum transduction. Assume we wish to transduce an
input microwave state with covariance matrix $\mathbf{V}_{\mathrm{in}}$ into
the optical domain. Following the standard continuous-variable teleportation
protocol \cite{unconditionalquantumteleportation,quantumteleportationwith,advancesinquantum}, the input microwave mode is mixed on a $50:50$ beam splitter with the
microwave component of the entangled resource $\mathbf{V}_{\mathrm{oe}}$.
Homodyne measurements are then performed on the two beam-splitter outputs to
obtain the conjugate quadratures $q$ and $p$. The measurement outcomes are
classically communicated to the optical node, where appropriate conditional
displacements are applied to the optical mode, and under ideal conditions
this recovers the input state on the optical side.

Consequently, the microwave input covariance $\mathbf{V}_{\mathrm{in}}$ is
transformed as $\mathbf{T\mathbf{V}_{\mathrm{in}}T}^{\mathrm{T}}+\mathbf{N}$
with%
\begin{equation}
\mathbf{T}=\kappa \mathbf{I}_{2}\text{,}  \label{e33}
\end{equation}%
\begin{eqnarray}
\mathbf{N} &=&\mathbf{V}_{\mathrm{A}}-\mathbf{V}_{\mathrm{C}}\mathbf{Z}_{%
\mathrm{2}}\mathbf{T-}(\mathbf{V}_{\mathrm{C}}\mathbf{Z}_{\mathrm{2}}\mathbf{%
T})^{\mathrm{T}}+\mathbf{T}^{\mathrm{T}}\mathbf{Z}_{\mathrm{2}}\mathbf{V}_{%
\mathrm{B}}\mathbf{Z}_{\mathrm{2}}\mathbf{T}  \notag \\
&=&(v\kappa ^{2}+u-2w\kappa )\mathbf{I}_{2}\text{,}  \label{e34}
\end{eqnarray}%
where $\kappa $ is an arbitrary classical feedforward gain applied to the
optical mode (For a detailed derivation, see Ref. \cite{microwaveandoptical}).
This defines a single-mode Bosonic channel and we can denote it as $\mathcal{%
N}^{\prime }(\eta ^{\prime },n_{e}^{\prime })$, with an effective thermal
noise $n_{e}^{\prime }$ and effective transmissivity%
\begin{equation}
\eta ^{\prime }=\kappa ^{2}\text{.}  \label{e35}
\end{equation}%
Specifically, if the gain factor is chosen as $\kappa <1$ the resulting
channel is a thermal loss channel, whereas for $\kappa >1$, it becomes a
thermal amplification channel. The corresponding added noise is given by%
\begin{equation}
n_{e}^{\prime }=\frac{v\kappa ^{2}+u-2w\kappa }{2|1-\kappa ^{2}|}-\frac{1}{2}%
\text{.}  \label{e36}
\end{equation}%
When the modification constant is $\kappa =1$, it represents a random
displacement channel with noise variance%
\begin{equation}
\sigma ^{2}=v+u-2w\text{.}  \label{e37}
\end{equation}

For a channel to reliably transmit quantum information, it must have a
positive quantum capacity. Again, we use the quantum capacity lower bound to
quantify the performance of the teleportation-induced transduction channel,
and the expression is given by \cite{gaussianquantuminformation}
\begin{equation}
Q_{\mathrm{LB}}^{\mathcal{N}^{\prime }}=%
\begin{cases}
\max \Bigl\{0,\log _{2}\bigl|\frac{\eta ^{\prime }}{1-\eta ^{\prime }}\bigr|%
-g(n_{e}^{\prime })\Bigr\}\text{,} & \eta ^{\prime }\neq 1\text{,} \\[8pt]
\max \Bigl\{0,\log _{2}\Bigl(\frac{2}{e\sigma ^{2}}\Bigr)\Bigr\}\text{,} &
\eta ^{\prime }=1\text{,}%
\end{cases}
\label{e38}
\end{equation}%
where the bound is tight when $n_{e}^{\prime }=0$. In the following text we
will select the optimal feedforward gain $\kappa $ for each parameter set
and use it to present and analyze the transduction performance.

\subsection{Performance of the transduction Channel}

In this subsection we present a detailed demonstration of the advantages
offered by EQT when the entangled resource exploits correlated noise. As
shown in Figs. \ref{f11}(a) and \ref{f11}(b), the effective added noise of
the teleportation-induced transduction channel $n_{e}^{\prime }$ can be
strongly suppressed over a specific range of coupling phase $\phi $. For
several degrees of noise correlation $\chi $, the noise can be reduced to
less than half of its value without correlations near the optimal phase $%
\phi =\pi /2$. The phase dependence of this noise suppression mirrors that
of the MO entanglement shown in Fig. \ref{f10}, with stronger entanglement
generally corresponding to lower effective added noise in the EQT channel.
Reducing the effective added noise is essential for single-photon level
transduction, and the enhancement of MO entanglement by noise correlations
may therefore substantially relax the stringent cryogenic requirements of
EQT.

According to Eq. (\ref{e38}), we plot the quantum capacity lower bound $Q_{\mathrm{LB}}^{\mathcal{N}^{\prime }}$ as a function of $\phi $ with different values of
$\chi $, as shown in Figs. \ref{f11}(c) and \ref{f11}(d). The presence of
noise correlations produces a pronounced enhancement of $Q_{\mathrm{LB}}^{%
\mathcal{N}^{\prime}} $ across a wide range of $\phi $. Moreover, as the
bath temperature increases, parameter regimes emerge near $\phi =\pi /2$ in
which $Q_{\mathrm{LB}}^{\mathcal{N}^{\prime }}=0$ for independent noise ($%
\chi =0$) but becomes positive when noise correlations are present ($\chi >0$%
), as illustrated in Fig. 3(d).

\begin{figure*}[tbh]
\centering \includegraphics[width=15cm]{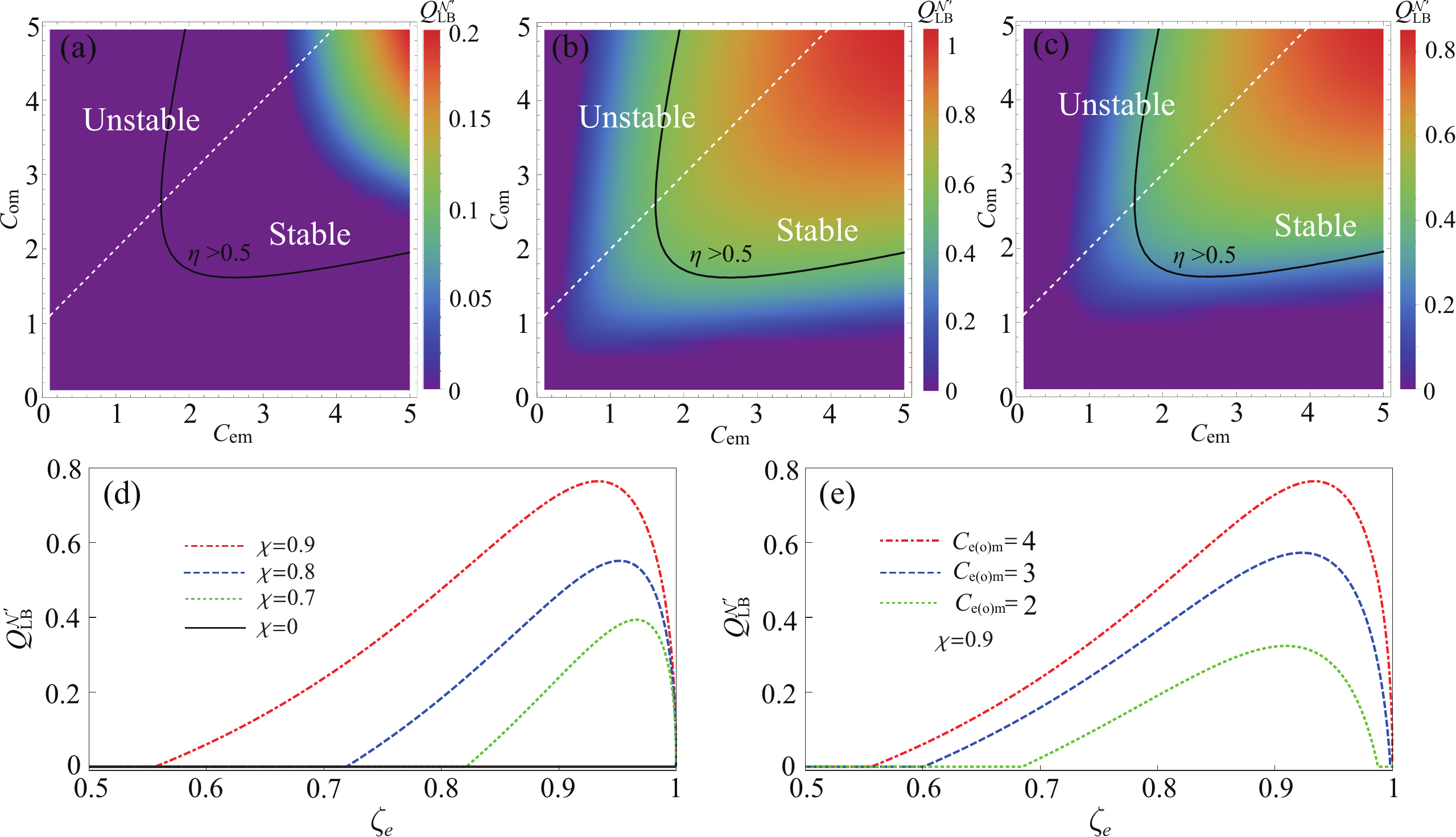}
\caption{Quantum capacity lower bound $Q_{\mathrm{LB}}^{\mathcal{N}^{\prime }}$ versus system
cooperativities $C_{\mathrm{om}}$ and $C_{\mathrm{em}}$ in panels (a)-(c),
and microwave extraction ratio $\protect\zeta _{\mathrm{e}}$ in panels (d)
and (e). The white dashed line separates the stable region in the lower right from the unstable region in the upper left. The black solid curve, which marks $\eta=1/2$, bounds the upper right parameter region in which DQT can potentially have a positive quantum capacity. The thermal noise is chosen as $n_{\mathrm{th}}=0.6$ in (a) and
(b), and $n_{\mathrm{th}}=1.6$ in (c)--(e). The system cooperativities are chosen
as $C_{\mathrm{om}}=$ $C_{\mathrm{em}}=4$ in panel (d) with different
degrees of correlation $\protect\chi =\{0,0.7,0.8,0.9\}$, and $\protect\chi
$ is fixed to $0.9$ in panel (e) with different cooperativities $\{C_{%
\mathrm{om}},C_{\mathrm{em}}\}=\{2,2\}$, $\{3,3\}$ and $\{4,4\}$. The
electromechanical coupling phase is chosen as $\protect\phi =\frac{\protect%
\pi }{2}$ for all plots. The optimal gain constant $\protect\kappa $ was
chosen for each parameter set. The other parameters are chosen to be the
same as in Fig. \protect\ref{f10}.}
\label{f12}
\end{figure*}

We further analyze the quantum capacity lower bound in the cooperativity
parameter $C_{\mathrm{om}}-C_{\mathrm{em}}$ space, as shown in Figs. \ref%
{f12}(a)-\ref{f12}(c). Specifically, Fig. \ref{f12}(a) corresponds to the
independent noise model, while Figs. \ref{f12}(b) and \ref{f12}(c)
correspond to the correlated noise model with $\chi =0.9$. Because the
generation of the MO entanglement resource relies on an optomechanical
two-mode squeezing interaction, an excessively strong blue-detuned optical
pump can cause the optomechanical parametric gain to exceed the effective
damping rate, thereby driving the system into a dynamically unstable regime
\cite%
{robustphotonentanglement,routh-hurwitzcriterionin,bipartiteandtripartite}.
Consequently, part of the parameter space is excluded by the stability
requirement. The white dashed line denotes the corresponding stability
boundary determined by the Langevin equation in Eq. (\ref{e26}), with the
physically relevant stable region lying below the boundary. Moreover, the
black curve marks the condition $\eta =1/2$ of the corresponding DQT
channel, and the upper right region enclosed by this curve gives the
potential positive capacity regime of a pure loss DQT channel. As shown in
Fig. \ref{f12}(a), without noise correlations, EQT can achieve a positive $Q_{\mathrm{LB}}^{\mathcal{N}^{\prime }}$ only within a limited part of the
stable parameter space. In contrast, Fig. \ref{f12}(b) shows that, for the
same local noise occupations, the correlated noise model substantially
enlarges the region with $Q_{\mathrm{LB}}^{\mathcal{N}^{\prime }}>0$ and
also increases its maximum value. Moreover, the positive capacity region
extends well beyond the potential positive regime of the pure loss DQT
channel indicated by the black curve. When the bath temperature increases,
the region of positive quantum capacity slightly shrinks, yet a substantial
area persists, as shown in Fig. \ref{f12}(c). This indicates that the
channel's transduction performance is relatively robust against thermal
noise.

Similar to the correlated noise model for DQT in Ref. \cite%
{correlatednoisecan}, the transduction capacity displays a nontrivial
dependence on the microwave extraction ratio $\zeta _{\mathrm{e}}$.
Specifically, the effect of $\zeta _{\mathrm{e}}$ to the transduction
channel is twofold. On the one hand, increasing $\zeta _{\mathrm{e}}$ enhances
the coupling of the microwave mode to the output port and reduces its
intrinsic loss channel in the MO entanglement generation process. This is
beneficial for extracting the microwave component of the entangled state and
suppressing the direct contribution of intrinsic loss. On the other hand, a
larger $\zeta _{\mathrm{e}}$ also means that less microwave intrinsic noise
participates in the contribution of correlated noise, and the advantage
associated with noise correlations is gradually weakened. As a result of
this trade-off, the dependence of $Q_{\mathrm{LB}}^{\mathcal{N}^{\prime }}$
on $\zeta _{\mathrm{e}}$ is nonmonotonic and an optimal $\zeta _{\mathrm{e}}$
exists that maximizes the lower bound. Fig. \ref{f12}(d) shows this behavior
for fixed system cooperativities and several values of $\chi $. As $\zeta _{%
\mathrm{e}}$ increases, $Q_{\mathrm{LB}}^{\mathcal{N}^{\prime }}$ rises from
zero to a peak, and then falls off rapidly. A similar trend appears in Fig. %
\ref{f12}(e), which plots $Q_{\mathrm{LB}}^{\mathcal{N}^{\prime }}$ versus $%
\zeta _{\mathrm{e}}$ for a fixed $\chi $ under several different
cooperativities. In the limit $\zeta _{\mathrm{e}}=1$, the microwave
intrinsic loss port is effectively removed, and the correlated noise model
reduces to the corresponding independent noise behavior with $\chi =0$. It
should be noted that in Fig. \ref{f12}(d) the value of $\zeta _{\mathrm{e}}$
that maximizes $Q_{\mathrm{LB}}^{\mathcal{N}^{\prime }}$ shifts toward
smaller with increasing $\chi $, whereas it shifts toward larger values as
the system cooperativities increase in Fig. \ref{f12}(e). Therefore,
optimizing $\zeta _{\mathrm{e}}$ is crucial for maximizing transduction
performance in experiments that exploit the correlated noise model.

\section{A Possible Mechanism for Engineering Noise Correlations}

In the preceding sections, the correlations between the intrinsic noise
inputs were characterized phenomenologically by the parameter $\chi $. This
description allowed us to evaluate their effects on the performance of both
DQT and EQT. Having established the advantages of the correlated noise
model, we now discuss a possible physical mechanism for generating the
required noise correlations in an EOM system. For the sake of simplicity, we
focus on modes $\hat{b}$ and $\hat{c}$. To generate intermode correlations
from initially independent thermal noise inputs, we introduce a two-mode squeezing electromechanical interaction. Specifically, this can be realized by driving the microwave mode with a strong coherent pump tone at frequency $\omega _{\mathrm{4}}$, blue-detuned from the microwave resonance by $\Delta _{\mathrm{4}}=\omega _{\mathrm{4}}-\omega _{\mathrm{e}}=\omega _{\mathrm{m}}$. The resulting Hamiltonian can be written in the rotating
frame as
\begin{equation}
\frac{\hat{H}^{\prime }}{\hbar }=g(\hat{b}\hat{c}+\hat{b}^{\dag }\hat{c}%
^{\dag })\text{,}  \label{e2}
\end{equation}%
where $g$ is the effective two-mode squeezing coupling rate.

As follows from the input-output relations derived earlier, the two-mode
squeezing interaction between $\hat{b}$ and $\hat{c}$ naturally produces
nonzero phase-sensitive cross moments between the mechanical and microwave
output fields, such as $\langle \hat{b}_{\mathrm{in}}\hat{c}_{\mathrm{in,i}%
}\rangle $. However, for the transduction process, the intrinsic noise
contributions from the mechanical and microwave loss channels are scattered
into the output fields through linear beam-splitter-type conversion
pathways. In this case, destructive interference between their noise
contributions is conveniently controlled by the phase-insensitive cross
moments, as represented by the nonzero second moments introduced in Eqs. (\ref{ex})
and (\ref{ex1}). Therefore, to exploit the generated correlations as a resource for
quantum transduction, it is necessary to tailor the structure of the noise
correlations.

\begin{figure*}[tbh]
\centering
\includegraphics[width=17.9cm]{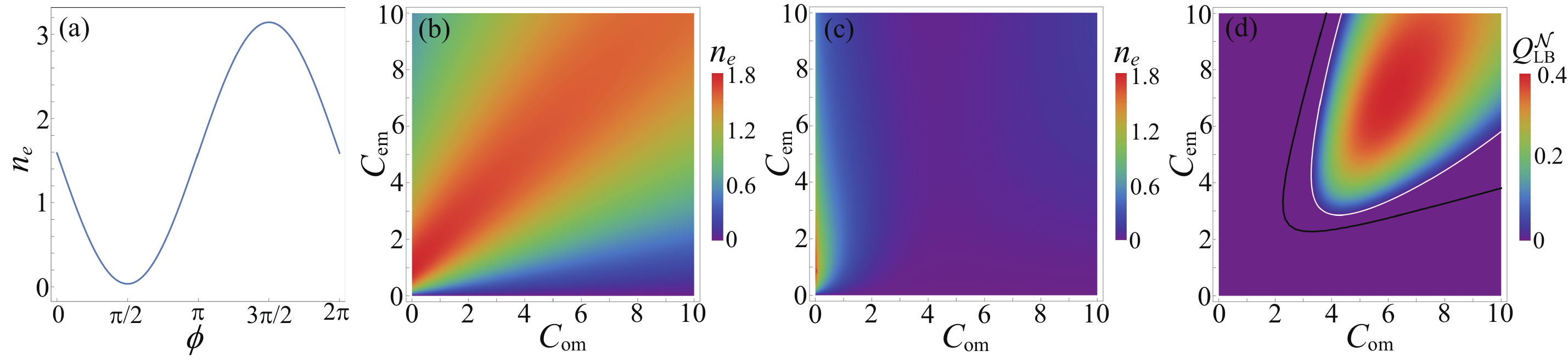}
\caption{Performance of direct quantum transduction. (a) Effective channel
noise $n_{e}$ as a function of the coupling phase $\protect%
\phi $. (b) and (c) Distributions of $n_{e}$ as functions of the
cooperativities $C_{\mathrm{om}}$ and $C_{\mathrm{em}}$ for $\protect\phi =0$
and $\protect\phi =\protect\pi /2$, respectively, with the same local noise $n_{b} =2.02$ and $n_{c} =2.29$. At $\protect\phi =0$ and $\pi$, the noise
correlations make no contribution, whereas $\protect\phi =\protect\pi /2$
gives the optimal destructive interference. (d) Quantum capacity lower bound
$Q_{\mathrm{LB}}^{\mathcal{N}}$ in the same cooperativity space at $\protect%
\phi =\protect\pi /2$. The black curve marks $\protect\eta =1/2$, separating
the regime with zero quantum capacity for direct transduction in the lower
left from the potentially positive capacity regime in the upper right. The
white curve bounds the upper right region in which $Q_{\mathrm{LB}}^{%
\mathcal{N}}>0$. For all plots, the microwave mode and optical mode
extraction ratios are $\protect\zeta _{\mathrm{e}}=0.8$ and $\protect\zeta _{%
\mathrm{o}}=0.9$, respectively. The other parameters are chosen as $C_{g}=C_{\protect\nu }=0.1$ and $%
\protect\theta =0$.}
\label{f2}
\end{figure*}

We then further reshape the correlation structure by applying a parametric
drive with frequency $\omega _{\mathrm{p}}$ to the microwave mode in
Hamiltonian (\ref{e2}). Such a single-mode squeezing operation can be implemented,
for example, in a superconducting microwave resonator with an intrinsically
tunable inductance, as commonly used in cavity-based Josephson parametric
amplifiers. For a resonant parametric drive with $\omega _{\mathrm{p}%
}=2\omega _{\mathrm{e}}$, the total Hamiltonian used to generate the
correlated noise can then be written in the rotating frame as%
\begin{equation}
\frac{\hat{H}_{3}}{\hbar }=g(\hat{b}\hat{c}+\hat{b}^{\dag }\hat{c}^{\dag
})+\nu (e^{-i\theta }\hat{c}^{\dag 2}+e^{i\theta }\hat{c}^{2})\text{,}
\label{e3}
\end{equation}%
where $\nu $ and $\theta $ are the strength and phase of the parametric
pump, respectively, and the microwave mode is described in a frame rotating
at $\omega _{\mathrm{p}}/2$. For convenience, we define the cooperativity $%
C_{g}=4g^{2}/\kappa _{\mathrm{m}}\kappa _{\mathrm{e}}$ and dimensionless
parameter $C_{\nu }=4\nu ^{2}/\kappa _{\mathrm{e}}^{2}$ to characterize the
relative strengths of the electromechanical interaction and the parametric
drive. In the following, unless otherwise specified, we restrict our
discussion to the stable regime when $C_{g}+2\sqrt{C_{\nu }}<1$.

\begin{figure*}[tbh]
\centering
\includegraphics[width=17.9cm]{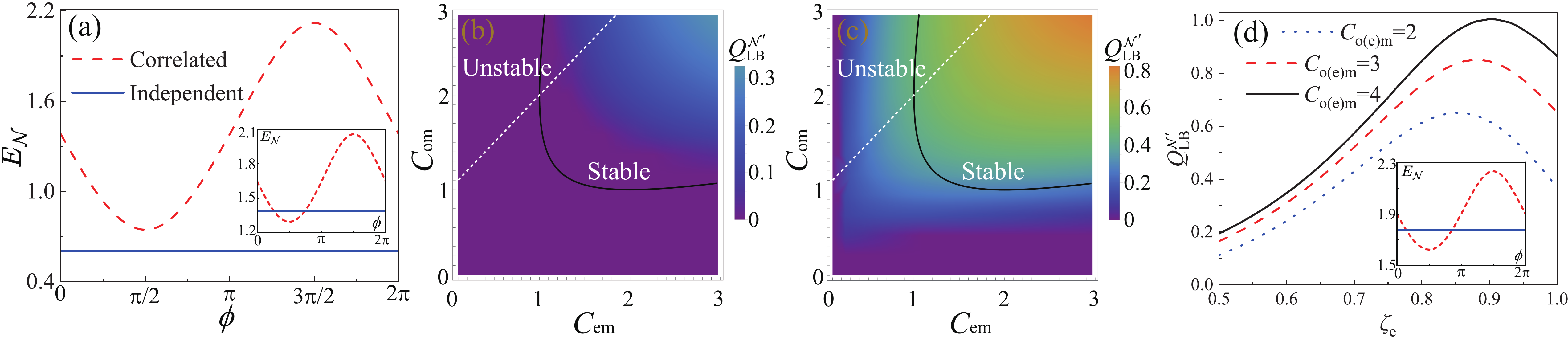}
\caption{Performance of entanglement-based quantum transduction. (a) Logarithmic negativity $E_{\mathcal{N}}$ of the MO entanglement resource as a function of the coupling phase $\phi$. The blue solid and red dashed curves represent independent and correlated noise, respectively. For the correlated-noise curves, the main panel, inset in (a), and inset in (d) use $\{C_g,C_\nu\}=\{0.1,0.1\}$, $\{0.1,0.05\}$, and $\{0.05,0.05\}$, corresponding to $\{n_b,n_c\}=\{2.02,1.14\}$, $\{0.89,0.30\}$, and $\{0.37,0.20\}$, respectively. Each independent-noise curve has the same local occupations $n_b$ and $n_c$ as its correlated-noise counterpart. (b) and (c) Quantum capacity lower bound $Q_{\mathrm{LB}}^{\mathcal{N}^{\prime}}$ over the system cooperativity space for independent and correlated noise, respectively. Both panels use $n_b=0.37$, $n_c=0.20$, and $\phi=3\pi/2$ is chosen for (c). The white dashed and black solid boundaries are defined as in Fig. \ref{f12}(a)--(c). (d) Quantum capacity lower bound $Q_{\mathrm{LB}}^{\mathcal{N}^{\prime}}$ as a function of the microwave coupling ratio $\zeta_e$. The blue dotted, red dashed, and black solid curves correspond to $\{C_{%
\mathrm{om}},C_{\mathrm{em}}\}=\{2,2\}$, $\{3,3\}$ and $\{4,4\}$, respectively, with $\phi=3\pi/2$. The correlated-noise results in panels (c) and (d) use $C_g=C_\nu=0.05$. The feedforward gain $\kappa$ is optimized for each parameter set. The other parameters are chosen as $\zeta_o=\zeta_e=0.9$ and $\theta=0$.}
\label{f3}
\end{figure*}

To determine the output noise correlations quantitatively, we now derive the
scattering matrix of the system. As before, we begin with the
Heisenberg--Langevin equations for each mode and the corresponding standard
input--output relations, which are given by%
\begin{equation}
\mathbf{\dot{a}}=\mathbf{Pa+Qa}_{\mathrm{in}}\text{,}  \label{e4}
\end{equation}%
\begin{equation}
\mathbf{a}_{\mathrm{out}}=\mathbf{P}^{\mathrm{T}}\mathbf{a-a}_{\mathrm{in}}%
\text{,}  \label{e5}
\end{equation}%
where the operators are grouped into the vectors $\mathbf{a}=(\hat{b},\hat{b}%
^{\dag },\hat{c},\hat{c}^{\dag })^{\mathrm{T}}$, $\mathbf{a}_{\mathrm{in}}=(%
\hat{b}_{\mathrm{in}},\hat{b}_{\mathrm{in}}^{\dag },\hat{c}_{\mathrm{in,c}},%
\hat{c}_{\mathrm{in,c}}^{\dag },\hat{c}_{\mathrm{in,i}},\hat{c}_{\mathrm{in,i%
}}^{\dag })^{\mathrm{T}}$ and $\mathbf{a}_{\mathrm{out}}=(\hat{b}_{\mathrm{%
out}},\hat{b}_{\mathrm{out}}^{\dag },\hat{c}_{\mathrm{out,c}},\hat{c}_{%
\mathrm{out,c}}^{\dag },\hat{c}_{\mathrm{out,i}},\hat{c}_{\mathrm{out,i}%
}^{\dag })^{\mathrm{T}}$. Here, the coefficient matrices are given by%
\begin{equation}
\mathbf{P}=\left(
\begin{array}{cccc}
-\frac{\kappa _{\mathrm{m}}}{2} & 0 & 0 & -ig \\
0 & -\frac{\kappa _{\mathrm{m}}}{2} & ig & 0 \\
0 & -ig & -\frac{\kappa _{\mathrm{e}}}{2} & -2i\nu e^{-i\theta } \\
ig & 0 & 2i\nu e^{i\theta } & -\frac{\kappa _{\mathrm{e}}}{2}%
\end{array}%
\right) \text{,}  \label{e6}
\end{equation}%
and%
\begin{equation}
\mathbf{Q}=\left(
\begin{array}{cccccc}
\sqrt{\kappa _{\mathrm{m}}} & 0 & 0 & 0 & 0 & 0 \\
0 & \sqrt{\kappa _{\mathrm{m}}} & 0 & 0 & 0 & 0 \\
0 & 0 & \sqrt{\kappa _{\mathrm{e,c}}} & 0 & \sqrt{\kappa _{\mathrm{e,i}}} & 0
\\
0 & 0 & 0 & \sqrt{\kappa _{\mathrm{e,c}}} & 0 & \sqrt{\kappa _{\mathrm{e,i}}}%
\end{array}%
\right) \text{.}  \label{e7}
\end{equation}%
The scattering relation is solved in the narrow-bandwidth limit as%
\begin{equation}
\mathbf{a}_{\mathrm{out}}=\mathbf{S}\cdot \mathbf{a}_{\mathrm{in}}\text{,}
\label{e8}
\end{equation}%
where $\mathbf{S}=\mathbf{Q}^{\mathrm{T}}(-\mathbf{P)}^{-1}\mathbf{Q}-%
\mathbf{I}_{6}$ is the scattering matrix.

The statistical properties of the input reservoirs are fully characterized
by the second-order moments of the input noise operators, and we define the
input noise correlation matrix through%
\begin{equation}
\mathbf{W}_{\mathrm{in}}=\langle \mathbf{a}_{\mathrm{in}}\mathbf{a}_{\mathrm{%
in}}^{\mathrm{T}}\rangle \text{.}  \label{e9}
\end{equation}%
Here, the coupling port of the microwave mode is taken to be in the vacuum
state, while the intrinsic loss ports of the mechanical and microwave modes
are assumed to be coupled to independent thermal reservoirs at the same
temperature $\mathcal{T}$. Therefore, $\mathbf{W}_{\mathrm{in}}$ is then
given by
\begin{equation}
\mathbf{W}_{\mathrm{in}}
=
\begin{pmatrix}
0 & 1+n_{\mathrm{th}}\\
n_{\mathrm{th}} & 0
\end{pmatrix}
\oplus
\begin{pmatrix}
0 & 1\\
0 & 0
\end{pmatrix}
\oplus
\begin{pmatrix}
0 & 1+n_{\mathrm{th}}\\
n_{\mathrm{th}} & 0
\end{pmatrix},
\label{e10}
\end{equation}
with $n_{\mathrm{th}}=\langle \hat{b}_{\mathrm{in}}^{\dag }\hat{b}_{\mathrm{%
in}}\rangle =\langle \hat{c}_{\mathrm{in,i}}^{\dag }\hat{c}_{\mathrm{in,i}%
}\rangle $. Since the input reservoirs are initially independent, all cross
correlations between different input ports vanish. From Eq. (\ref{e8}), the
output noise correlation matrix can be obtained as%
\begin{equation}
\mathbf{W}_{\mathrm{out}}=\langle \mathbf{a}_{\mathrm{out}}\mathbf{a}_{%
\mathrm{out}}^{\mathrm{T}}\rangle =\mathbf{S\mathbf{W}_{\mathrm{in}}S}^{%
\mathrm{T}}\text{.}  \label{e11}
\end{equation}%
In $\mathbf{W}_{\mathrm{out}}$, we define the local noise occupations as $%
n_{b}=\langle \hat{b}_{\mathrm{out}}^{\dag }\hat{b}_{\mathrm{out}}\rangle $
and $n_{c}=\langle \hat{c}_{\mathrm{out,i}}^{\dag }\hat{c}_{\mathrm{out,i}%
}\rangle $. $n_{b}$ is determined by $C_g$, $C_\nu$ and $\mathcal{T}$, whereas $n_{c}$ is additionally depends on $\zeta_{\mathrm e}$. Importantly, the output fields from the intrinsic
loss ports of the mechanical and microwave modes acquire nonzero intermode
correlations, such as the phase-insensitive cross moment $\langle \hat{b}_{%
\mathrm{out}}^{\dag }\hat{c}_{\mathrm{out,i}}\rangle $ and the corresponding
conjugate moment.

Analogous to the environmental memory effect in bosonic channels \cite{restoringquantumcommunication}, these correlations remain appreciable for a finite interval and can influence the subsequent transduction process. During this interval, the output fields serve as correlated noise inputs to the intrinsic dissipation channels, allowing their effects on transduction to be controlled through the relevant scattering pathways by tuning the system parameters. Under the Markov approximation, the intermode correlations decay exponentially and vanish in the long-time limit, with the memory lifetime governed primarily by the effective decoherence rates of the generated mechanical and microwave noise fields. In the following analysis, we use the initial values of the correlated second moments.

We now apply the noise correlations generated through this mechanism to both
DQT and EQT and examine their effects on the corresponding transduction
performance. For DQT, the contributions from the noise correlations vanish at $\phi=0$ and $\pi$, while reduce the effective channel noise $n_e$ throughout the range $0<\phi<\pi$ for the parameters considered here, as shown in Fig. \ref{f2}(a). Figs. \ref{f2}(b) and \ref{f2}%
(c) further compare the distributions of $n_{e}$ in the system cooperativity
space at $\phi =0$ and $\phi =\pi /2$, respectively, with the same local
noise occupations $n_{b}$ and $n_{c}$. At $\phi =\pi /2$, destructive interference
reduces $n_{e}$ close to zero over a broad region of the cooperativity
space, as shown in Fig. \ref{f2}(c). Moreover, the region with $Q_{\mathrm{LB%
}}^{\mathcal{N}}>0$ nearly covers the entire parameter regime in which a
pure loss DQT channel can support a positive quantum capacity, as shown in
Fig. \ref{f2}(d). In contrast, if the intrinsic noise inputs are
uncorrelated, no region with positive quantum capacity can be obtained.

For EQT, Fig. \ref{f3}(a) shows the logarithmic negativity $E_{\mathcal{N}}$
of the MO entanglement as a function of $\phi $. The influence of the engineered noise correlations can be understood in two distinct processes. First, the mechanical mode acts as an effective collective channel that transfers the input noise correlations to the MO state. This becomes evident upon adiabatically eliminating $\hat b$, after which the same noise operator $\hat b_{\mathrm{in}}$ drives both $\hat a^{\dagger}$ and $\hat c$, thereby promoting MO entanglement. Second, $\phi$ controls the interference between the mechanically mediated $\hat b_{\mathrm{in}}$ contribution and the direct $\hat c_{\mathrm{in,i}}$ contribution to the microwave output. This phase-dependent interference is distinct from the collective channel effect. By contrast, the phenomenological $\chi$ model (\ref{ex1}) contains only phase-insensitive cross moments and lacks the additional phase-sensitive correlations produced by this mechanism, so correlated noise can either enhance or degrade the MO entanglement, as shown in Fig. \ref{f10}. For the parameters of the main panel in Fig. \ref{f3}(a), the collective channel enhancement outweighs unfavorable interference throughout the phase range, keeping $E_{\mathcal{N}}$ above the independent-noise result and producing a maximum at $\phi=3\pi/2$. However, the two insets in panels (a) and (d) show that for other $n_b$ and $n_c$ values unfavorable interference can dominate near the minimum and reduce $E_{\mathcal{N}}$ below the independent-noise value.

Figs. \ref{f3}(b) and \ref{f3}(c) show the distributions of the quantum
capacity lower bound $Q_{\mathrm{LB}}^{\mathcal{N}^{\prime }}$ of the
teleportation-induced transduction channel in the system cooperativity space
for independent and correlated noise, respectively. Consistent with the
results in Figs. \ref{f12}(a) and \ref{f12}(b), correlated noise markedly
improves the transduction performance compared with independent noise at the
same local noise occupations. In Fig. \ref{f3}(d), we examine the dependence
of $Q_{\mathrm{LB}}^{\mathcal{N}^{\prime }}$ on the microwave extraction
ratio $\zeta _{\mathrm{e}}$ for several sets of system cooperativities. The
resulting trends are consistent with those shown in Fig. \ref{f12}(e). For
the parameters considered, the optimal values of $\zeta _{\mathrm{e}}$ lie
between $0.8$ and $0.9$.

\section{Conclusion}

In this work, we analyzed
in detail how noise correlations can improve quantum transduction
performance. Specifically, we considered correlations between the microwave and mechanical noise inputs and showed that they can be exploited to suppress the transduction channel noise when the relevant system parameters are appropriately tuned. For DQT, the terms arising from the noise
correlations can interfere destructively with the independent noise
contributions, thereby directly reducing the effective channel noise. This
reduction substantially relaxes the stringent conditions required for DQT to
achieve positive quantum capacity.

For EQT, the noise correlations can affect the quality of the generated MO
entanglement. This influence is directly reflected in the changes
to the covariance matrix elements of the MO output state. By jointly
optimizing the relevant system parameters, the entanglement of the MO state can
be substantially enhanced compared with that generated under independent
noise. When employed in teleportation, the improved entanglement resource
markedly reduces the effective added noise of the teleportation-induced
transduction channel. Consequently, correlated noise substantially increases
the quantum capacity lower bound of EQT. In the system cooperativity space,
the region with positive quantum capacity is considerably larger than that
obtained under independent noise at the same bath temperature. In addition, we examined the nonmonotonic dependence of
the EQT quantum capacity on the microwave extraction ratio. This behavior
should be carefully considered when selecting operating parameters for
optimal transduction performance.

Finally, we discussed a potential physical mechanism for generating the
required noise correlations. By tuning the
electromechanical coupling and applying a suitable parametric drive, nonzero
cross moments can be established between the mechanical and microwave output noise fields. We
then applied the generated correlations to both DQT and EQT and evaluated
their effects on the transduction performance for representative system
parameters. These results demonstrate the improvements achievable under
appropriate operating conditions and provide theoretical guidance for future
experimental realization.

More broadly, because the underlying mechanism relies on interference between correlated noise inputs, the resulting noise reduction is not limited to a specific platform and applicable to any hybrid transduction platforms with multiple dissipative channels. This
approach may substantially relax the cryogenic requirements imposed on
practical transducers. Combining the correlated noise model with complementary
cooling methods, such as radiative cooling \cite%
{radiativecoolingof,cavitypiezo-mechanicsfor,quantumentanglementbetween,radiativecoolingofa,bench-topcoolingof}%
, could further reduce the residual channel noise and facilitate quantum
transduction at the single-photon level.

\begin{acknowledgements}
We acknowledge the funding support from Xi'an Jiaotong University (Grant No. 11301224010717), Shaanxi Fundamental Science Research Project for Mathematics and Physics (Grant No. 25JSY006), and the Youth Scientist funding support from Shaanxi Province (Grant No. 2024SYJ21).
\end{acknowledgements}

\appendix

\section{CALCULATION OF THE LOGARITHMIC NEGATIVITY WITH CORRELATED NOISE}

\label{aa}

In the appendix, we derive the logarithmic negativity $E_{\mathcal{N}}$ of the output MO Gaussian state with noise correlation (\ref{ex1}). The covariance matrix of the MO output takes the form in Eq. (\ref{e32}), and the matrix elements are given by
\begin{eqnarray}
u &=&1+\frac{8C_{\mathrm{om}}\zeta _{\mathrm{o}}}{(1+C_{\mathrm{em}}-C_{%
\mathrm{om}})^{2}}\cdot  \notag \\
&&\Big[\big(1+n_{\mathrm{th}}+C_{\mathrm{em}}(1+n_{\mathrm{th}}-n_{\mathrm{th%
}}\zeta _{\mathrm{e}})\big)  \notag \\
&&+2n_{\mathrm{th}}\sqrt{C_{\mathrm{em}}(1-\zeta _{\mathrm{e}})}\cdot \chi
\sin \theta \Big]\text{,}
\end{eqnarray}
\begin{eqnarray}
v &=&1+\frac{8}{(1+C_{\mathrm{em}}-C_{\mathrm{om}})^{2}\zeta _{\mathrm{e}%
}^{-1}}\cdot  \notag \\
&&\Big[C_{\mathrm{em}}(C_{\mathrm{om}}+n_{\mathrm{th}})-(C_{\mathrm{om}%
}-1)^{2}(\zeta _{\mathrm{e}}-1)n_{\mathrm{th}}  \notag \\
&&+2(C_{\mathrm{om}}-1)n_{\mathrm{th}}\sqrt{C_{\mathrm{em}}(1-\zeta _{%
\mathrm{e}})}\cdot \chi \sin \theta \Big]\text{,}
\end{eqnarray}
and
\begin{equation}
w=\sqrt{c^{2}+d^{2}}\text{,}
\end{equation}
where $c$ and $d$ are the elements in the cross-covariance block before
applying local phase-space rotations
\begin{eqnarray}
c &=&\frac{4\sqrt{C_{\mathrm{em}}C_{\mathrm{om}}\zeta _{\mathrm{e}}\zeta _{%
\mathrm{o}}}\cos \theta }{(1+C_{\mathrm{em}}-C_{\mathrm{om}})^{2}}\cdot
\notag \\
&&\Big[1+C_{\mathrm{em}}+C_{\mathrm{om}}+2C_{\mathrm{om}}n_{\mathrm{th}%
}-2(C_{\mathrm{om}}-1)n_{\mathrm{th}}\zeta _{\mathrm{e}}  \notag \\
&&+4n_{\mathrm{th}}\sqrt{C_{\mathrm{em}}(1-\zeta _{\mathrm{e}})}\cdot \chi
\sin \theta \Big]\text{,}
\end{eqnarray}
\begin{eqnarray}
d &=&-\frac{4\sqrt{C_{\mathrm{om}}\zeta _{\mathrm{e}}\zeta _{\mathrm{o}}}}{%
(1+C_{\mathrm{em}}-C_{\mathrm{om}})^{2}\sqrt{C_{\mathrm{em}}}}\cdot  \notag
\\
&&\Big[2n_{\mathrm{th}}\sqrt{C_{\mathrm{em}}(1-\zeta _{\mathrm{e}})}(C_{%
\mathrm{om}}-1-C_{\mathrm{em}}\cos 2\theta )\cdot \chi  \notag \\
&&+C_{\mathrm{em}}\Big(1+C_{\mathrm{em}}+C_{\mathrm{om}}  \notag \\
&&+2n_{\mathrm{th}}\big(1+(1-\zeta _{\mathrm{e}})C_{\mathrm{om}}\big)\Big)%
\sin \theta \Big]\text{.}
\end{eqnarray}

To evaluate the logarithmic negativity, we perform partial transposition with respect to the microwave mode. In phase space, this operation corresponds to reversing the sign of the microwave momentum quadrature
\begin{equation}
\hat{p}_{\mathrm{e}}\longrightarrow-\hat{p}_{\mathrm{e}}\text{.}
\end{equation}
The partially transposed covariance matrix is therefore given by
\begin{equation}
\widetilde{V}_{\mathrm{oe}}=\Lambda V_{\mathrm{oe}}\Lambda\text{,}
\end{equation}
where $\Lambda=I_2\oplus Z_2$, and we obtain
\begin{equation}
\widetilde{V}_{\mathrm{oe}}
=
\begin{pmatrix}
uI_2 & wI_2\\
wI_2 & vI_2
\end{pmatrix}\text{,}
\end{equation}
The symplectic eigenvalues of $\widetilde{V}_{\mathrm{oe}}$ are determined from the positive eigenvalues of $\left|i\Omega\widetilde{V}_{\mathrm{oe}}\right|$, where
\begin{equation}
\Omega=J\oplus J\text{,}
\end{equation}
and
\begin{equation}
J=
\begin{pmatrix}
0&1\\
-1&0
\end{pmatrix}\text{.}
\end{equation}
For a general two-mode Gaussian state, the smaller symplectic eigenvalue of the partially transposed covariance matrix is
\begin{equation}
\widetilde{\nu}_{-}
=
\sqrt{
\frac{
\widetilde{\Delta}
-
\sqrt{\widetilde{\Delta}^{2}-4\det V_{\mathrm{oe}}}
}{2}
}\text{,}
\end{equation}
where
\begin{equation}
\widetilde{\Delta}=\det V_\mathrm{A}+\det V_\mathrm{B}-2\det V_\mathrm{C}\text{.}
\end{equation}
For the covariance matrix considered here, we have $\det V_\mathrm{A}=u^2$, $\det V_\mathrm{B}=v^2$, $\det V_\mathrm{C}=-w^2$ and $\det V_{\mathrm{oe}}=(uv-w^2)^2$. The output MO state is entangled when
\begin{equation}
\widetilde{\nu}_{-}<1\text{.}
\end{equation}
Therefore, the logarithmic negativity is given by
\begin{equation}
E_{\mathcal N}
=
\max\left\{
0,-\log_2\widetilde{\nu}_{-}
\right\}\text{.}
\end{equation}

\bibliography{sub.bib}

\end{document}